\documentclass[12pt]{iopart}
\usepackage{graphicx}

\newcommand{\bra}[1]{\ensuremath{\left\langle #1\right|}}
\newcommand{\ket}[1]{\ensuremath{\left|#1\right\rangle}}

\begin{document}

\title[Entangling microscopic defects via a macroscopic quantum shuttle]{Entangling microscopic defects via a macroscopic quantum shuttle}

\author{G~J~Grabovskij$^1$, P~Bushev$^1$, J~H~Cole$^{2,3}$, C~M\"uller$^{4,3}$, J~Lisenfeld$^1$, A~Lukashenko$^1$ and A~V~Ustinov$^{1,3}$}
\address{$^{1}$ Physikalisches Institut, Karlsruhe Institute of Technology, D-76128 Karlsruhe, Germany}
\address{$^{2}$ Institut f\"ur Theoretische Festk\"{o}rperphysik, Karlsruhe Institute of Technology, D-76128 Karlsruhe, Germany}
\address{$^{3}$ DFG-Center for Functional Nanostructures (CFN), D-76128 Karlsruhe, Germany}
\address{$^{4}$ Institut f\"{u}r Theorie der Kondensierten Materie, Karlsruhe Institute of Technology, D-76128 Karlsruhe, Germany}
\ead{ustinov@kit.edu}

\begin{abstract}
In the microscopic world, multipartite entanglement has been achieved with various types of nanometer sized two-level systems such as trapped ions, atoms and photons. On the macroscopic scale ranging from micrometers to millimeters, recent experiments have demonstrated bipartite and tripartite entanglement for electronic quantum circuits with superconducting Josephson junctions. It remains challenging to bridge these largely different length scales by constructing hybrid quantum systems. Doing this may allow for manipulating the entanglement of individual microscopic objects separated by macroscopically large distances in a quantum circuit. Here we report on the experimental demonstration of induced coherent interaction between two intrinsic two-level states (TLSs) formed by atomic-scale defects in a solid via a superconducting phase qubit. The tunable superconducting circuit serves as a shuttle communicating quantum information between the two microscopic TLSs. We present a detailed comparison between experiment and theory and find excellent agreement over a wide range of parameters. We then use the theoretical model to study the creation and movement of entanglement between the three components of the quantum system.
\end{abstract}

\vspace{2pc}
%\noindent{\it Keywords}: superconducting qubits, Josephson junctions, two-level states, microwave spectroscopy, entanglement
\submitto{\NJP}
\maketitle

\section{Introduction}
Controllable coherent interaction between individual quantum systems is one of the fundamental prerequisites for quantum information processing. This interaction allows one to selectively entangle systems and to transfer quantum information between its individual parts. The demonstration of such tunable interaction, allowing each individual part of a system to controllably interact with every other part, is therefore a major step towards demonstrating the possible use of a particular architecture for quantum information processing~\cite{Nielsen:2000}. In atomic systems such quantum information transfer and processing is established e.g., via long range Coulomb~\cite{Roos:2004} or magnetic dipole interactions~\cite{Neumann:2008}. Photonic systems utilize the Kerr-type interaction in non-linear crystals~\cite{Knill:2001} while for superconducting qubits the interaction is defined by the circuit design~\cite{Makhlin:2001,Devoret:2004}. In order to achieve tunability, it has proven useful to introduce an additional quantum system acting as a mediator of the interaction between two parts. This role can be played e.g. by a resonant cavity or additional, so called \emph{ancilla} qubits~\cite{Plastina:2003,Silinpaa:2008}. The use of ancilla qubits has been demonstrated e.g. with electron and nuclear spins of $^{13}$C atoms in NV-defect centers of diamond~\cite{Neumann:2008} and with superconducting flux qubits~\cite{Niskanen:2007}. For superconducting systems the use of a microwave cavity placed on the same chip has proven very useful~\cite{Fink:2009, Ansmann:2009}. For superconducting qubits, due to their nature as part of an electronic circuit, the effect of the environment on the dynamics is often very strong. This has led to a much better understanding of the nature of the environment as well as to improved qubit designs which are insensitive to certain of its characteristics~\cite{Koch:2007,Manucharyan:2009}. One part of the environment is formed by so-called two-level defect states (TLSs). Ensembles of these TLSs are a general model for decoherence~\cite{Phillips:1972,Anderson:1972} in a wide variety of systems including glasses and mechanical resonators~\cite{Zolfagharkhani:2005,MacFarlane:2006,Arcizet:2009,Venkatesan:2010,Macha:2010,Hoehne:2010}. In superconducting phase qubits, which have large-area Josephson junctions, one often finds signatures of individual two-level systems resonantly interacting with the qubit~\cite{Simmonds:2004}. TLSs are in general thought to be detrimental to the operation of the qubit~\cite{Martinis:2005, Palomaki:2010}, but since they are often more coherent than the qubit and their potential use as a quantum memory has been demonstrated~\cite{Neeley:2008}, it has been proposed to use the TLS itself as computational qubits~\cite{Zagoskin:06}.

In this paper, we demonstrate a coherent interaction between two microscopic defect states mediated by a superconducting phase qubit. The TLSs have fixed but different resonant frequencies, and the phase qubit works as a frequency-tunable shuttle communicating quantum information between them. The observed dynamical quantum beating signal between all three systems verifies the multipartite interaction and the basic operations presented here offer the possibility to establish coherent control over many TLSs coupled to the Josephson junction of any flux or phase qubit. A sketch of the experimental setup is shown in figure 1(a). The sample is maintained at a temperature of around $35$ mK in a dilution refrigerator. For details on the experimental setup we refer the reader to Ref. \cite{Lisenfeld:2010}.
The coupling between the qubit and TLSs leads to characteristic anticrossings in qubit spectroscopy (figure 1(b)). From their positions and sizes one can infer the level splitting of the TLSs as well as the strength of their coupling to the qubit. Using resonant microwave driving of the TLSs when the qubit is far detuned~\cite{Lisenfeld:2010:Temp}, we are able to determine their coherence properties. Table 1 gives eigenenergies $\omega$, coupling strengths $v$ as well as the relaxation and dephasing rate for the qubit and both TLSs.
Note that the decoherence times of the TLSs are much longer in comparison to the phase qubit, and one of them shows $T_{1}$-limited dephasing \cite{Lisenfeld:2010:Temp}.

\begin{figure} %Fig. 1
\begin{center}
 \includegraphics[width=\textwidth]{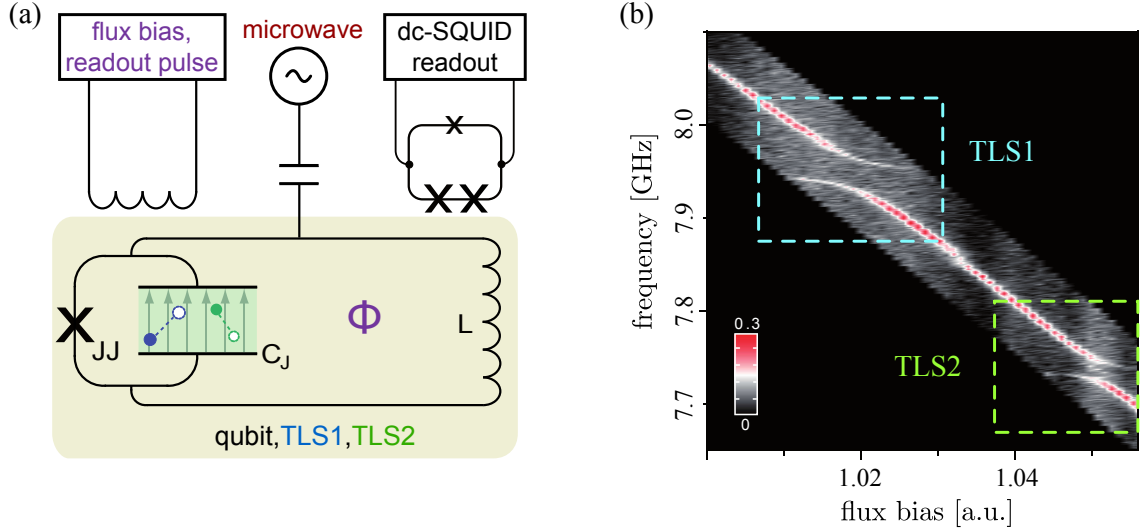}
\caption{(a) Sketch of the experimental setup and circuit of the phase qubit with two TLSs residing inside the qubit's Josephson junction. (b) Excitation spectrum of the phase qubit showing two avoided level crossings due to the coupling with TLS1 and TLS2. The color bar denotes excitation probability of the qubit.}
\end{center}
\end{figure}

~\\

\begin{table}
\begin{center}
	\begin{tabular}{c|c|c|c}
                    & Qubit  & \, TLS1 \, & \, TLS2 \, \\
		\hline
		$\omega/2\pi$ [GHz] & -     & 7.946 & 7.735 \\
		$v/2\pi$ [MHz]      & -      & 36    & 23 \\
		$T_1$ [ns]          & 120    & 380   & 410 \\
		$T_2$ [ns]          & 90     & 580   & 810 \\
	\end{tabular}
\end{center}
\caption{The characteristics of the three subsystems: qubit, TLS1 and TLS2. The resonance frequencies and the couplings between the qubit and the TLSs are denoted by $\omega$ and $v$ respectively. The range for the holding position of the qubit $\omega_h$ varies over 7.65 - 8.00 GHz, covering the resonance frequencies of the TLSs. The characteristic times are denoted as $T_1$ (relaxation time) and $T_2$ (dephasing time as measured use a Ramsey experiment).}
\end{table}

\section{Establishing a coherent interaction between the three subsystems}

The pulse sequence applied to the circuit %to establish a coherent interaction between all three subsystems 
is shown in Fig 2. Initially, the entire system is prepared in its ground state. The qubit is detuned away from both TLSs and excited with a $\pi$-pulse. A fast flux bias pulse (of rise time 2 ns) brings it in resonance with TLS2 and keeps it there for a fixed time to perform an $\sqrt{\rm{iSWAP}}$-gate~\cite{Neeley:2008,Schuch:2003}. After this gate, half of excitation remains in the qubit and half is transferred to TLS2, resulting in an entangled state between these subsystems. To induce an interaction between all three components of our system, the qubit is then tuned to a frequency $\omega_h$ and is held there for a time $t_h$. This hold position ($\omega_h$) is varied over a range which includes the resonance frequencies of both TLSs. Depending on the detuning between qubit and TLS1 or TLS2 ($\Delta_{q1}$, $\Delta_{q2}$) and the hold time, the qubit acquires a phase with respect to TLS2 as well as exchanging population with one or both TLSs.  The population and acquired phase of the qubit can be revealed by performing an additional $\sqrt{\rm{iSWAP}}$-gate between the qubit and TLS2 followed by a readout of the qubit. The results can be compared with an interference pattern. Depending on the relative phase between the qubit and TLS2 the interference can be constructive or destructive, resulting in energy being transferred to the qubit or to TLS2. The advantage of the protocol used here, in comparison to e.g., just measuring the beating between qubit and a TLS in dependence on detuning, is that the visibility does not decrease with detuning but depends only on the phase difference between the qubit and TLS2 (and the dephasing processes).

\begin{figure} %Fig. 2
\begin{center}
 \includegraphics[width=0.6\textwidth]{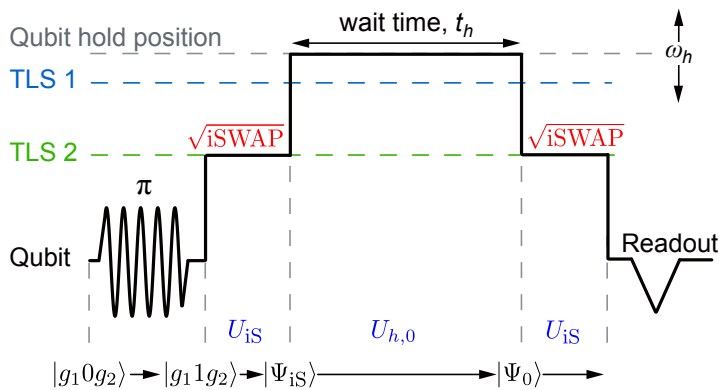}
\caption{The experimental sequence used to establish a coherent interaction between a phase qubit and a pair of TLSs. The qubit is first excited using a $\pi$-pulse. The frequency of the qubit is then tuned to be resonant with TLS2 in order to realise an $\sqrt{\rm{iSWAP}}$-gate. The qubit is then biased at some frequency $\omega_h$ lying in the range that covers the resonances of both TLSs and hold there for some time $t_h$. A subsequent $\sqrt{\rm{iSWAP}}$ with TLS2 and readout then gives a direct signal of the interaction between the TLSs induced by the qubit.}
\end{center}
\end{figure}

\section{Analysis of the experimental results}

The measured escape probability, which corresponds to the probability to measure the qubit in its excited state, is presented on figure 3(a). The experiment was performed in two ranges such that the qubit is held close to the resonance with either TLS, resulting in two chevron patterns. figure 3(c) shows the result of a numerical simulation, which corresponds well to the measured results. The time dependence of the system is calculated via the evolution of the density matrix of the whole system~\cite{Bushev:2010} including decoherence in Lindblad form~\cite{Lindblad:1976}. The parameters for the simulation are taken from independent measurements of the various coupling parameters and decoherence rates (cf. Table 1). To study the system dynamics in more detail, we solve the coherent evolution of the system analytically at key points of interest.  The resulting expressions for the escape probability (see below) display the same qualitative behavior as the full numerical simulations shown in figure 3(c),(d) and figure 4.

\begin{figure} %Fig. 3
\begin{center}
 \includegraphics[width=\textwidth]{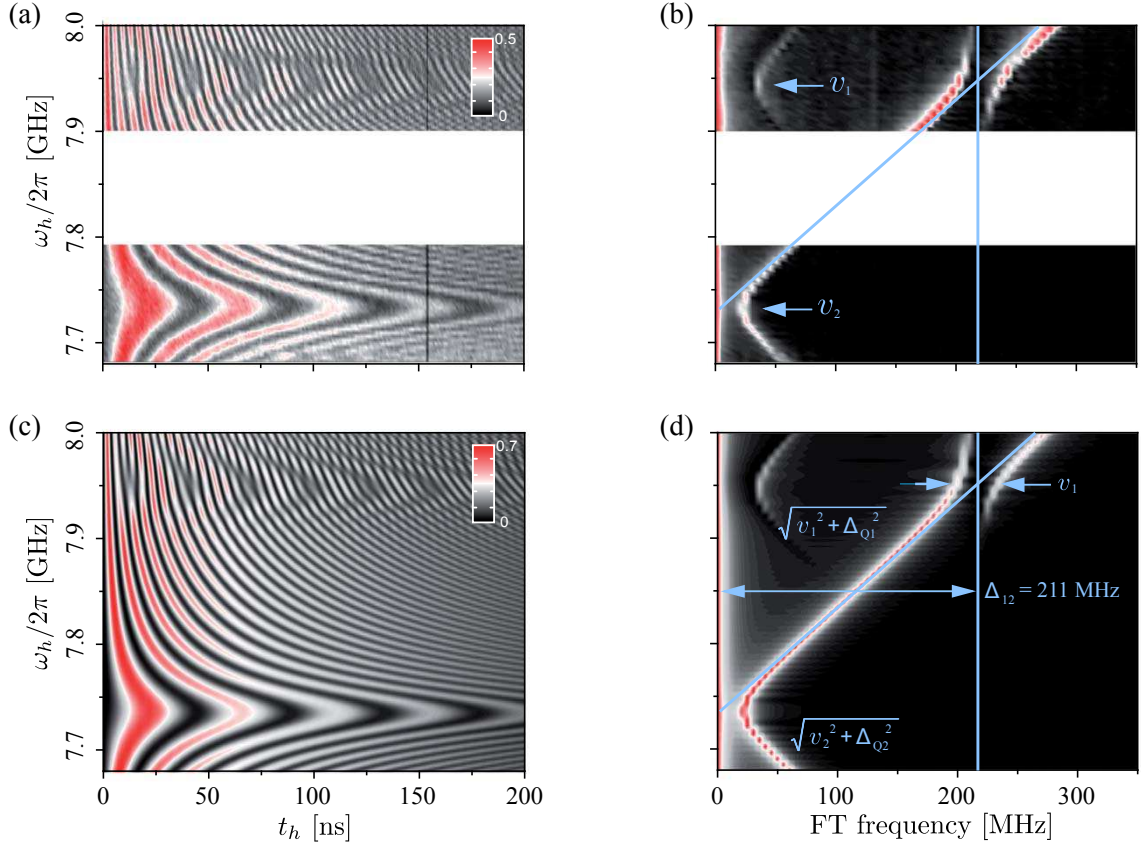}
\caption{Time and frequency domain evolution of the combined qubit-TLS1-TLS2 system. Experimental (a) and theoretical (c) beating signal of the qubit state with two TLSs. The colorbar shows the measured/calculated escape probability for the phase qubit. Experimental (b) and theoretical (d) Fourier transform of the beating signal showing hyperbolas due to interaction between the qubit and each of TLSs. The vertical blue line indicates the TLS1-TLS2 detuning $\Delta_{21}$. The diagonal blue line follows the relation $\omega_2+\Delta_{q2}$. The anticrossing on TLS2 hyperbola appears in resonance with TLS1 at a frequency of $\Delta_{12}$ and indicates the established interaction between two defects via the phase qubit.}
\end{center}
\end{figure}

\begin{figure} %Fig. 4
\begin{center}
 \includegraphics[width=0.6\textwidth]{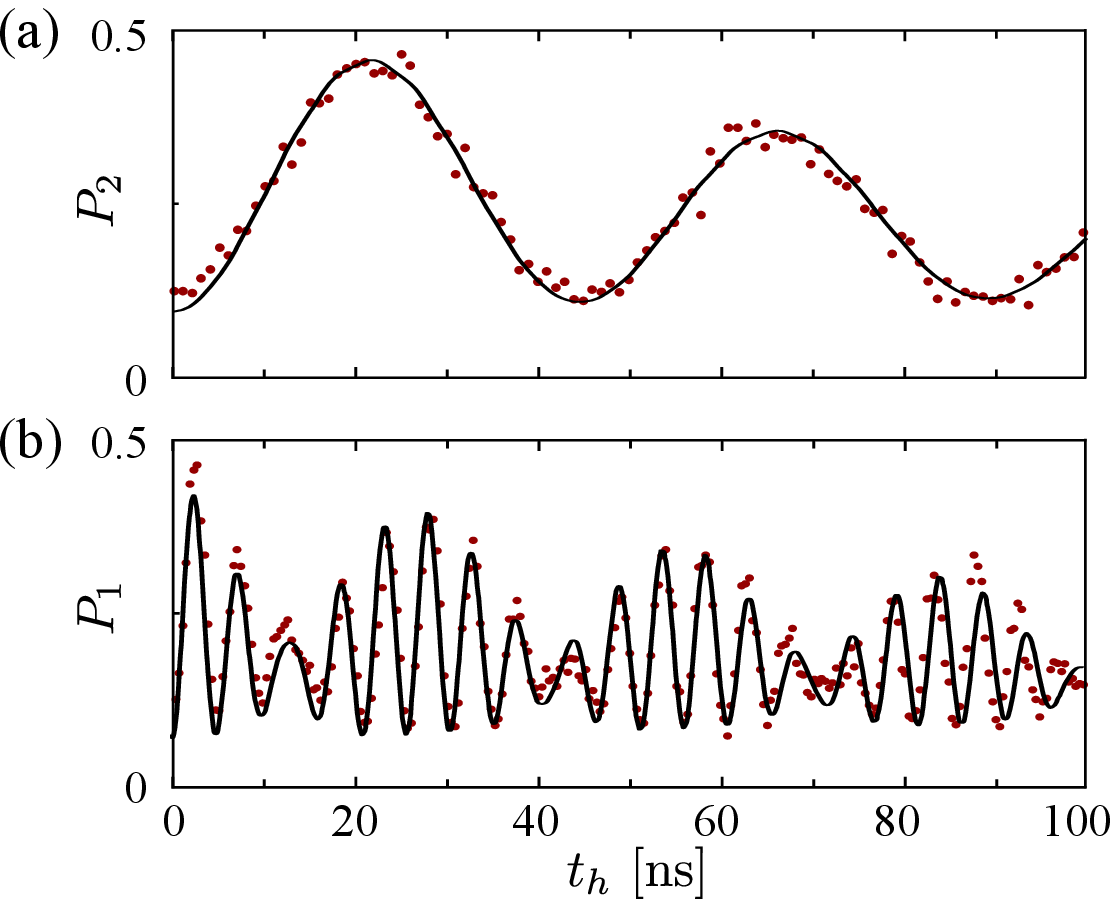}
\caption{Dynamics of the final qubit population at each of the TLS resonances. The simulated (solid line) and measured (dots) qubit occupation probability at the holding position of the qubit $\omega_h$ in resonance with TLS2 (a) and TLS1 (b).}
\end{center}
\end{figure}

The lower chevron in figure 3(a),(c) corresponds to the situation when the qubit is held close to TLS2 and the observed oscillations simply correspond to energy transfer between qubit and TLS2. If we neglect the coupling to TLS1, the probability of finding the qubit in its excited state while exactly in resonance with TLS2 can be expressed as
\begin{equation}
  	P_2(t_h)\approx\frac{1}{2}\left(1-\cos v_2t_h\right) \,.
  	\label{eq:pe2}
\end{equation}
This shows the simple oscillations with the frequency of the qubit-TLS coupling that enables one to coherently transfer information between qubit and TLS. The comparison between the theory and experiment for this case is shown in figure 4(a) at the point $\omega_h = \omega_2$. Both curves are extracted from the plots in Figs. 3(a) and (c). The probability of the theoretical curve is scaled linearly to match the observed measurement visibility.

As mentioned above, the protocol used here allows us to detune the qubit far from the resonance with TLS2 without loss of visibility. This can be seen in figure 3(a), where a clear oscillatory signal is observed over a range of more than $300$ MHz. The equivalent theoretical plot can be seen on figure 3(c). In the vicinity of TLS1, the chevron pattern develops an additional overlaid feature, which is similar in structure to that of TLS2. It is exactly this feature which is the signature of the induced TLS-TLS coupling. At the point where the qubit is held resonant with TLS1, we can obtain a simplified analytical expression for the measurement probability by assuming that the qubit is decoupled from one TLS when being in resonance with the other (the derivation of equations (\ref{eq:pe2}) and (\ref{eq:pe1}) will be discussed in detail later):
\begin{equation}
  	P_1(t_h) \approx
  	\left(\frac38+\frac18\cos v_1 t_h-\frac12\cos\frac{v_1t_h}2\cos\Delta_{21}t_h\right)\,,
  	\label{eq:pe1}
\end{equation}
where $\Delta_{21} = 2\pi \times 211$ MHz is the frequency difference between the two TLS. This expression shows temporal beating of the qubit population due to the induced coherent interaction with the three frequencies $v_1$, $\Delta_{21}-v_1/2$ and $\Delta_{21}+v_1/2$. The experimental and theoretical curves at $\omega_h=\omega_1$ are extracted from the Figs. 3(a) and (c) and plotted in figure 4(b) (the theoretical curve is again scaled linearly to match the experiment).

To see the three frequency components more clearly, we show the Fourier transform (FT) of the measured, figure 3(b), and simulated, figure 3(d), temporal evolution of the escape probability. The frequency spectra contain the hyperbola $\sqrt{v_{2}^2+\Delta_{q2}^2}$ which relates the swap frequency and the qubit-TLS2 detuning $\Delta_{q2}$. This hyperbola shows an anticrossing in the vicinity of TLS1, which has a splitting size equal to $v_1$ = 36 MHz at the detuning frequency of $\Delta_{21}$ = 211 MHz. Furthermore, an additional hyperbola $\sqrt{v_{1}^2+\Delta_{q1}^2}$ appears due to the direct coupling  between qubit and TLS1. Here we stress that the appearing of the \emph{three fundamental frequencies} in the system's dynamic indicates the established interaction between all three parts of the system, and can not be attributed to the interaction between any two parts. 

\section{Theoretical description}

In addition to the numerical simulation of the three coupled 2-level quantum systems including decoherence effects~\cite{Bushev:2010}, we also present a simplified theoretical picture yielding the two analytical expressions (\ref{eq:pe2}) and (\ref{eq:pe1}). The Hamiltonian $H$ of the entire system depends on the frequency of the qubit $\omega_q$ 
and of the TLSs $\omega_i,\ i=1,2$ and the interaction parts between the qubit and the TLSs $v_i,\ i=1,2$:
\begin{eqnarray}
  	H_0(\omega_q)=\frac\hbar2\left[\omega_q\sigma_z+\omega_1\tau_{1,z}+\omega_2\tau_{2,z}+\right. \nonumber\\ \qquad\qquad    
  	     \left. v_1(\sigma_-\tau_{1,+}+\sigma_+\tau_{1,-}) + v_2(\sigma_-\tau_{2,+}+\sigma_+\tau_{2,-})\right], 
\end{eqnarray}
with $\sigma$ and $\tau_i,\ i=1,2$ being the Pauli matrices for the qubit and TLSs, respectively. In the following, $\ket{0},\ \ket{1}$ will denote the states of the qubit and $\ket{g_i},\ \ket{e_i},\ i=1,2$ that of TLS1 and TLS2, respectively. The exact solution of the time evolution of the system, while tractable, is too complicated to provide useful insight. A clearer understanding is obtained by solving the time evolution without decoherence in the limit that the qubit in the vicinity of one TLS is decoupled from the other TLS. This approximation is justified in the case that $\Delta_{21}\gg v_1, v_2$. In this limit, the Hamiltonian becomes
\begin{eqnarray}
  	H_i=H_0(\omega_q=\omega_i)\approx \nonumber \\
  	\frac\hbar2\left[\omega_i\sigma_z+\omega_1\tau_{1,z}+\omega_2\tau_{2,z} + v_i\left(\sigma_-\tau_{i,+}+\sigma_+\tau_{i,-}\right)\right],
\end{eqnarray}
where $i=1,2$ indicates resonance with TLS1 or TLS2, respectively. We are still able to describe the beating due to the effect of the two TLSs as the qubit is brought into resonance with each in turn. The operator $U_{\rm{iS}}=\exp(-iH_2t_{\rm{iS}}/\hbar)$ describes the $\sqrt{\rm{iSWAP}}$-gate operation between the qubit and TLS2 while neglecting the interaction with TLS1. Here $t_{\rm{iS}}= \pi/(2v_2)$ is the time needed for the gate. The evolution of the state vector during the holding time is given by the unitary operator $U_{h,0}(\omega_h,t_h)=\exp(-iH_0(\omega_h)t_h/\hbar)$, which under our approximation gives us the approximate operators $U_{h,i}(t_h)=\exp(-iH_it_h/\hbar),\ i=1,2$.

Starting with the ground state $\ket{g_1 0 g_2}$, the qubit is excited which results in the state $\ket{g_1 1 g_2}$. The state after the first $\sqrt{\rm{iSWAP}}$ is then given by $\ket{\Psi_{\rm{iS}}}=U_{\rm{iS}}\ket{g_1 1 g_2}=(\ket{g_1 1 g_2}-i\ket{g_1 0 e_2})/\sqrt{2}$. In the next step, the interaction between all components of our system is established. This is described by the equation $\ket{\Psi_0(\omega_h,t_h)}=U_{h,0}(\omega_h,t_h)U_{\rm{iS}}\ket{g_1 1 g_2}$. Finally, the measured result for the escape probabilities $P_{1,2}$ in equations (\ref{eq:pe2}) and (\ref{eq:pe1}) are then calculated via
\begin{equation}
  	P_i(t)\approx\left|\bra{1}U_{\rm{iS}}U_{h,i}(t_h)U_{\rm{iS}}\ket{g_1 1 g_2}\right|^2,\ i=1,2.
\end{equation}

\section{Three-way entanglement between qubit and TLSs}
However, for theoretical analysis of the tripartite dynamics the interesting state of the system is just before the final $\sqrt{\rm{iSWAP}}$, as this is before TLS2 is disentangled from the system. At the resonance frequency of TLS1 $\omega_h=\omega_1$ we find the approximated expression for the state vector to be
\begin{eqnarray}
 	|\Psi_0(\omega_h=\omega_1,t_h)\rangle\approx U_{h,1}(t_h)U_{\rm{iS}}\ket{g_1 1 g_2} \nonumber \\
 	= -\frac{i}{\sqrt{2}}\ket{g_1 0 e_2}
 		+\frac{e^{-i \Delta_{21} t_h}}{\sqrt{2}} \left(-i\sin \frac{v_1 t_h}{2}\ket{e_1 0 g_2} + \cos \frac{v_1 t_h}{2}\ket{g_1 1 g_2}\right).
 	\label{eq:psi_t}
\end{eqnarray}
As expected, half of the population is located in TLS2 and its state $\ket{g_10e_2}$ does not show any coherent time evolution. In contrast, the qubit and TLS1 exchange population while accumulating a phase with respect to TLS2 with the frequency $\Delta_{21}$. The population of the various components obtained from the numerical treatment are plotted in figure 5a. Here, fast oscillations in the probability curves of the qubit and TLS2, which have opposite phase, can be recognized. They indicate that the qubit-TLS2 coupling is still not negligible but does not disrupt the overall dynamics. The vanishing and reappearing of the fast oscillations indicate the presence of two frequencies which are close to each other. This is consistent with the spectrum of the oscillations shown in figure 3(b),(d), and by equation \ref{eq:pe1}, which yields the two frequencies $\Delta_{21}\pm v_1/2$.

\begin{figure} %Fig. 5
\begin{center}
 \includegraphics[width=0.6\textwidth]{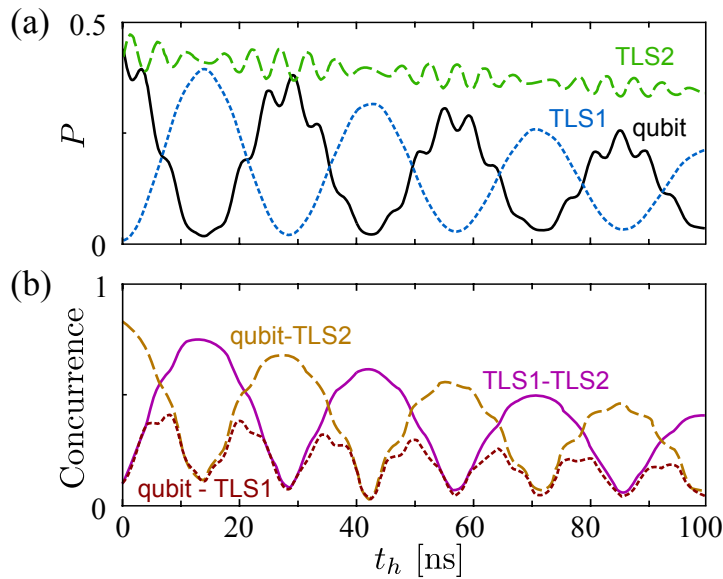}
\caption{Simulation of the excitation probability of, and entanglement between, each of the three system components before the second $\sqrt{\rm{iSWAP}}$. (a) Simulated excitation probabilities of the qubit (solid), TLS1 (dotted) and TLS2 (dashed), while holding the qubit at resonance with TLS1 for the time $t_h$. (b) The concurrence between each individual pair of subsystems (qubit-TLS2 - dashed line, qubit-TLS1 - dotted line, TLS1-TLS2 - solid line) after tracing out the third component.}
\end{center}
\end{figure}

To further study this interplay between the two defects, we use our theoretical model to calculate the entanglement present in the tripartite system. In figure 5(b) we plot the concurrence~\cite{Wootters:98} between each pair of system components (while tracing over the third). The concurrence is an entanglement measure yielding a number between 0 (no entanglement) and 1 (full entanglement). The initial entanglement between qubit and TLS2 (via the $\sqrt{\rm{iSWAP}}$) subsequently oscillates between the different components, reaching a maximum between TLS2 and either qubit or TLS1, depending of the location of the majority of the population. In contrast, the concurrence between qubit and TLS1 shows a beating with double frequency, as the maximal entanglement between these two subsystems is established twice a cycle.  Interspersed between these points, we see points of genuine tripartite entanglement, due to the interaction between all three components. This multipartite entanglement is of the $W$-state class, as our initial condition restricts us to the single-excitation subspace and therefore the `3-tangle' is precisely zero~\cite{Coffman:2000, Duer:2000}. We note that tripartite entanglement of the W-type was recently reported for systems consisting of three superconducting qubits \cite{Neeley:2010,DiCarlo:2010}.

\section{Summary}
In conclusion, we have presented evidence for controlled interaction between two microscopic defect states mediated by a phase qubit. During the implementation of the pulse sequence the tunable qubit serves as a quantum shuttle, providing a bridge between two TLSs and establishing \emph{a coherent tripartite interaction}. The Fourier spectrum of the observed qubit-TLSs beating contains clear evidence of such an interaction between all parts of the system over a time scale limited by the coherence of the system. This demonstration shows the ability to use coherent interaction between many TLSs for implementation of quantum gates.

\section*{Acknowledgements}
We would like to thank M. Ansmann and J. M. Martinis (UCSB) for providing us with the sample that we measured in this work. This work was supported by the CFN of DFG, the EU projects SOLID and MIDAS, and the U.S. ARO under Contract No. W911NF-09-1-0336.

\section*{References}

\bibliographystyle{unsrt}

\bibliography{references}

\begin{thebibliography}{10}

\bibitem{Nielsen:2000}
Nielsen~M A and Chuang~I L.
\newblock {\em {Q}uantum {C}omputation and {Q}uantum {I}nformation}.
\newblock Cambridge University Press, 2000.

\bibitem{Roos:2004}
Roos~C F, Riebe M, H\"{a}ffner H, H\"{a}nsel W, Benhelm J, Lancaster G~P T,
  Becher C, Schmidt-Kaler F, and Blatt R.
\newblock {\em Science}, 304:1478, 2004.

\bibitem{Neumann:2008}
Neumann P, Mizuochi N, Rempp F, Hemmer P, Watanabe H, Yamasaki S, Jacques V,
  Gaebel T, Jelezko F, and Wrachtrup J.
\newblock {\em Science}, 320:1326, 2008.

\bibitem{Knill:2001}
Knill E, Laflamme R, and Milburn~G J.
\newblock {\em Nature}, 409:46, 2001.

\bibitem{Makhlin:2001}
Makhlin Y, Sch\"on G, and Shnirman A.
\newblock {\em Rev. Mod. Phys.}, 73:357, 2001.

\bibitem{Devoret:2004}
Devoret~M H, Wallraff A, and Martinis~J M.
\newblock {\em ArXiv:cond-mat/0411174}, 2004.

\bibitem{Plastina:2003}
Plastina F and Falci G.
\newblock {\em Phys. Rev. B}, 67:224514, 2003.

\bibitem{Silinpaa:2008}
Silinp\"{aa}~M A, Park~J I, and Simmonds~R W.
\newblock {\em Nature}, 449:438, 2007.

\bibitem{Niskanen:2007}
Niskanen~A O, Harrabi K, Yoshihara F, Nakamura Y, Lloyd S, and Tsai~J S.
\newblock {\em Science}, 316:723, 2007.

\bibitem{Fink:2009}
Fink~J M, Bianchetti R, Baur M, Goeppl M, Steffen L, Filipp S, Leek~P J, Blais
  A, and Wallraff A.
\newblock {\em Phys. Rev. Lett}, 103:083601, 2009.

\bibitem{Ansmann:2009}
Ansmann M, Wang H, Bialczak~R C, Hofheinz M, Lucero E, Neeley M, O'Connell~A D,
  Sank D, Weides M, Wenner J, Cleland~A N, and Martinis~J M.
\newblock {\em Nature}, 461:504, 2009.

\bibitem{Koch:2007}
Koch J, Yu~T M, Gambetta J, Houck~A A, Schuster~D I, Majer J, Blais A,
  Devoret~M H, Girvin~S M, and Schoelkopf~R J.
\newblock {\em Phys. Rev. A}, 76:042319, 2007.

\bibitem{Manucharyan:2009}
Manucharyan~V E, Koch J, Glazman~L I, and Devoret~M H.
\newblock {\em Science}, 326:113, 2009.

\bibitem{Phillips:1972}
Phillips~W A.
\newblock {\em J. Low Temp. Phys.}, 7:351, 1972.

\bibitem{Anderson:1972}
Anderson~P W, Halperin~B I, and Varma~C M.
\newblock {\em Philosophical Magazine}, 25:1478, 1972.

\bibitem{Zolfagharkhani:2005}
Zolfagharkhani G, Gaidarzhy A, Shim S, Badzey~R L, and Mohanty P.
\newblock {\em Phys. Rev. B}, 72:224101, 2005.

\bibitem{MacFarlane:2006}
Macfarlane~R M, Sun Y, Sellin~P B, and Cone~R L.
\newblock {\em Phys. Rev. Lett}, 96:033602, 2006.

\bibitem{Arcizet:2009}
Arcizet O, Rivi{\`e}re R, Schliesser A, Anetsberger G, and Kippenberg~T J.
\newblock {\em Phys. Rev. A}, 80:021803, 2009.

\bibitem{Venkatesan:2010}
Venkatesan A, Lulla~K J, Patton~M J, Armour~A D, Mellor~C J, and
  Owers-Bradley~J R.
\newblock {\em J. Low Temp. Phys.}, 685:224101, 2010.

\bibitem{Macha:2010}
Macha P, van~der Ploeg S H~W, Oelsner G, Il'ichev E, Meyer H-G, W{\"u}nsch S,
  and Siegel M.
\newblock {\em Appl. Phys. Lett.}, 96:062503, 2010.

\bibitem{Hoehne:2010}
Hoehne F, Pashkin~Yu A, Astafiev O, Faoro L, Ioffe~L B, Nakamura Y, and Tsai~J
  S.
\newblock {\em Phys. Rev. B}, 81:184112, 2010.

\bibitem{Simmonds:2004}
Simmonds~R W, Lang~K M, Hite~D A, Nam S, Pappas~D P, and Martinis~J M.
\newblock {\em Phys. Rev. Lett.}, 93:7, 2004.

\bibitem{Martinis:2005}
Martinis~J M, Cooper~K B, McDermott R, Steffen M, Ansmann M, Osborn~K D, Cicak
  K, Seongshik O, Pappas D~P andSimmonds R~W, and Yu~C C.
\newblock {\em Phys. Rev. Lett.}, 95:210503, 2005.

\bibitem{Palomaki:2010}
Palomaki~T A, Dutta~S K, Lewis~R M, Przybysz~A J, Paik H, Cooper~B K, Kwon H,
  Anderson~J R, Lobb~C J, and Wellstood~F C.
\newblock {\em Phys. Rev. B}, 81:144503, 2010.

\bibitem{Neeley:2008}
Neely M, Ansmann M, Bialczak~R C, Hofheinz M, Katz N, Lucero E, O'Connell A,
  Wang H, Cleland A, and Martinis~J M.
\newblock {\em Nat. Phys.}, 4:523, 2008.

\bibitem{Zagoskin:06}
Zagoskin~A M, Ashhab S, Johansson~J R, and Nori F.
\newblock {\em Phys. Rev. Lett.}, 97:077001, 2006.

\bibitem{Lisenfeld:2010}
Lisenfeld J, M{\"u}ller C, Cole~J H, Bushev P, Lukashenko A, Shnirman A, and
  Ustinov~A V.
\newblock {\em Phys. Rev. B}, 81:100511(R), 2010.

\bibitem{Lisenfeld:2010:Temp}
Lisenfeld J, M{\"u}ller C, Cole~J H, Bushev P, Lukashenko A, Shnirman A, and
  Ustinov~A V.
\newblock {\em Phys. Rev. Lett.}, 105:230504, 2010.

\bibitem{Schuch:2003}
Schuch N and Siewert J.
\newblock {\em Phys. Rev. A}, 67:032301, 2003.

\bibitem{Bushev:2010}
Bushev P, M{\"u}ller C, Cole~J H, Lisenfeld J, Lukashenko A, Shnirman A, and
  Ustinov~A V.
\newblock {\em Phys. Rev. B}, 82:134530, 2010.

\bibitem{Lindblad:1976}
Lindblad G.
\newblock {\em Communications in Mathematical Physics}, 48:119, 1976.

\bibitem{Wootters:98}
Wootters~W K.
\newblock {\em Phys. Rev. Lett.}, 80:2245, 1998.

\bibitem{Coffman:2000}
Coffman V, Kundu J, and Wootters~W K.
\newblock {\em Phys. Rev. A}, 61:052306, 2000.

\bibitem{Duer:2000}
D\"ur W, Vidal G, and Cirac~J I.
\newblock {\em Phys. Rev. A}, 62:062314, 2000.

\bibitem{Neeley:2010}
Neeley M, Bialczak~R C, Lenander M, Lucero E, Mariantoni M, O'Connell~A D, Sank
  D, Wang H, Weides M, Wenner J, Yin Y, Yamamoto T, Cleland~A N, and Martinis~J
  M.
\newblock {\em Nature}, 467:570, 2010.

\bibitem{DiCarlo:2010}
DiCarlo L, Reed~M D, Sun L, Johnson~B R, Chow~J M, Gambetta~J M, Frunzio L,
  Girvin~S M, Devoret~M H, and Schoelkopf~R J.
\newblock {\em Nature}, 467:574, 2010.

\end{thebibliography}

\end{document}